# Ultra-low noise microwave extraction from fiber-based optical frequency comb.

J. Millo[1], R. Boudot[2], M. Lours[1], P. Y. Bourgeois[2], A. N. Luiten[3], Y. Le Coq[1], Y. Kersalé[2], and G. Santarelli[*1]

[1]*LNE-SYRTE, Observatoire de Paris, CNRS, UPMC, 61 Avenue de l'Observatoire, Paris, France*

[2]*FEMTO-ST Institute, CNRS and ENSMM, Besançon, France*

[3]*School of Physics, University of Western Australia, Crawley 6009, Australia*

[*]*Corresponding author: giorgio.santarelli@obspm.fr*

In this letter we report on an all optical-fiber approach to the generation of ultra-low noise microwave signals. We make use of two erbium fiber mode-locked lasers phase locked to a common ultra stable laser source to generate an 11.55 GHz signal with an unprecedented relative phase noise of -111 dBc/Hz at 1 Hz from the carrier. The residual frequency instability of the microwave signals derived from the two optical frequency combs is below $2.3 \times 10^{-16}$ at 1s and about $4 \times 10^{-19}$ at $6.5 \times 10^4$ s (in 5Hz Bandwidth, three days continuous operation).

OCIS codes: (140.3510) Lasers, fiber; (140.4050) Mode-locked lasers; (120.3930) Metrological *instrumentation; (230.0250) Optoelectronics*

Low phase-noise, microwave signals with high long term stability and reliability are of prime importance in a variety of scientific and technological fields, such as, for example, atomic frequency standards, radar and remote sensing, communications and navigation, high-speed electronics, very long baseline interferometry and high-precision timing distribution and synchronisation [1]. At present, the lowest noise microwave sources are based on ultra-low noise quartz oscillators or cryogenic sapphire oscillators [2]. However, the second of these sources is not commercially available and has high maintenance costs, while even the best quartz oscillator sources have a noise which is performance-limiting in some leading-edge applications. In contrast, over recent times the stabilization of lasers to high finesse optical cavities has become a mature technology and reliable devices based around vibration-immune cavities show a fractional frequency stability (FFS) that is routinely about $1 \times 10^{-15}$ from 0.1s to 100s or better [3-5]. At the same time, femtosecond laser optical frequency combs have emerged as the ultimate low noise optical-to-microwave frequency divider delivering the potential for low noise microwave generation [6]. By combining an ultra-stable laser with this frequency division technique it should be possible to generate a signal with a phase noise performance that surpasses all existing microwave sources over a broad range of the Fourier spectrum. In this letter we demonstrate the development of a highly reliable all-fiber optical system that demonstrates a low-jitter and low-phase noise microwave signal generation with a leading-edge performance consistent in magnitude over several days of measurement.

The pioneering work of S. Diddams and co-workers at NIST on Titanium-Sapphire-based Optical Frequency Comb (TSOFC), has demonstrated residual phase noise as low as - 110 dBc/Hz at 1Hz for a carrier frequency of 10 GHz as well as a frequency instability of $6.5 \times 10^{-16}$ at 1s [7, 8]. Moreover the same group has investigated the mechanisms of excess noise in microwave signal extraction from optical pulse train (see [9,10] and references therein). Despite these outstanding results the TSOFC suffers from a lack of repeatability in the magnitude of the phase noise as well as an insufficient long term operational reliability necessary for many applications. Fiber-based Optical Frequency Combs (FOFC), on the other hand, are suitable for continuous operation and have already demonstrated rather low phase noise signal generation for cesium atomic fountain clocks operation [11-13].

In this Letter we make use of 250 MHz repetition rate ($f_{rep}$) FOFCs that include *f*-2*f* interferometers for measuring the carrier-envelope offset frequency (Menlo Systems GmbH, M-Comb +P250+XPS1500). A mode of the FOFC is phase locked to an ultra-stable cw laser, which allows synthesis of microwave signals through photodetection of the light pulses (which provides $f_{rep}$ and its harmonics within the detector's bandwidth). Due to the phase lock loop, the repetition rate of the laser is phase-coherent with the optical reference signal but divided down in frequency by a large factor (~800000). It has the same FFS and time jitter as that of the optical reference, with a small unavoidable noise added by the division process. We investigate the impact of this added noise to the FFS as well as to the phase noise of the synthesized microwave signals. Our setup is described in Fig. 1: two quasi-identical FOFCs are independently phase-locked to a shared and common optical reference consisting of a cw fiber laser at 1542.14 nm stabilized onto an ultra stable cavity [14]. The setup is tuned such that both FOFCs have the same $f_{rep}$ (hence exhibiting exactly the same division factor from optical to microwave). Since the intrinsic noise of the optical reference is common mode, comparing the 2 microwave signals generated by the FOFCs completely characterizes the phase noise added by the optical to microwave division process.

The lock technique used to stabilize the repetition rate of the FOFC, shown on Figure 1, is an improved version of the one described in [13]. The whole arrangement makes use of fiber pigtailed single mode optical components. Critical components, in particular power splitters and Optical Add and Drop modules (OADM), were hand-selected to minimize phase noise using our experience from ref. [15]. The 30 mW output (100 fs duration pulses) from the mode-locked laser oscillator was sent through a polarisation controller to an OADM (three port interference optical filter centered at 1542.14 nm with bandwidth 0.8 nm). The spectrally narrow filtered output is combined with the CW light from the ultra stable laser (of optical frequency $\nu_{cw}$). The resulting beat-note signal $f_b=\nu_{cw}-Nxf_{rep}-f_0$ (where $N$ is a large integer and $f_0$ is the carrier envelope offset frequency) is detected on a *InGaAs* photodiode. From the *f*-2*f* built-in interferometer we obtain a signal $f_0$ which is mixed with $f_b$ and filtered to produce a frequency signal $\nu_{cw}-Nxf_{rep}$ independent of $f_0$. This signal is filtered by a tracking oscillator and then digitally divided by 128. Finally this divided signal is compared to a reference synthesized with a direct digital synthesizer (DDS1 see Fig. 1) to produce a phase error signal. This signal, processed in a simple analog loop filter, then controls the pump power of the femtosecond laser. The servo bandwidth is about 120 kHz, which allows robust and reliable phase locking to the reference cw laser. To generate a signal at the repetition rate (and its harmonics) the other output of the OADM, which contains nearly all of the mode-locked oscillator power (~10mW), is injected into a high-speed InGaAs pigtailed photodiode (Discovery model DSC40S, 20 GHz bandwidth). The second FOFC is locked with a similar technique but due to the lack of a fast pump power control port the bandwidth is limited to 20-30 kHz.

For both FOFC systems, the low level output signal (about -30 dBm) of the microwave extraction photodiode is filtered by a very low insertion losses microwave filter centered at 11.55 GHz and amplified using very low flicker phase noise amplifiers[*]. One of the outputs is further filtered and amplified up to a power of 5 dBm to saturate the LO port of a double balanced microwave mixer. The RF port is driven with -8 dBm from the other FOFC system.

Figure 2a shows the measured relative phase noise of the microwave signals at 11.55 GHz for two independent systems (black solid line) as well as the noise level of the readout system (photodetection,amplifiers and mixer) (dashed grey line). The detection floor was measured by driving both detection photo-diodes with the same FOFC. The residual phase noise of the amplifiers is slightly below that of the complete readout system. As can be seen on Figure 2a the measured noise is limited by the measurement system below ~100Hz Fourier frequencies. Nevertheless we achieve an very low phase noise level of about $-111-10\log(f)$ dBc/Hz for 1 Hz$<f<$100 Hz for the two systems. The measurement system noise could be reduced by utilizing cross-correlation and/or carrier suppression techniques (see ref. [10] and references therein). However our approach guarantees that this level of phase noise is available to any application with a classical phase detection system. The timing jitter integrated from 0.1 Hz to 100 kHz is about 1.8 fs rms.

For a valid comparison with previous results [7] we calculate the equivalent phase noise on a 10 GHz carrier, and express it for a single FOFC. The result is about $-115-10\log(f)$ dBc/ for 1 Hz$<f<$100 Hz Fourier frequency. This is, to our knowledge, the best result ever reported close to the carrier for microwave generation using optical frequency combs. It is also important to note the very low level of spurious peaks in the spectrum and that no post-data processing or filtering has been used to produce this result.

---

[*] AML*Communications* models 612L2201&812PNA2401 at the front end, 812PNB2401 boosting amplifier

Beyond 100 Hz the limited control bandwidth of one of the combs does not allow us to investigate the ultimate lower limit to the microwave generation process. However if a larger control bandwidth is achieved, using for example electro optical modulators inside the laser cavities, we would not expect any major impediment in reaching the limit set by the photodetection shot-noise (around -140 dBc/Hz) at Fourier frequencies higher than 100 Hz.

We have evaluated the FFS of the generated microwave signals by sampling the low pass filtered (5 Hz) voltage output of the microwave mixer. The top plot of figure 3 shows the relative timing jitter of the generated microwave signals over 72h. The bottom plot of the same figure shows FFS (overlapping Allan deviation) against the measurement time from a 74 hours dataset of continuous operation. The FFS for a single system is about $1.6 \times 10^{-16}$ (1 s-10 s) scaling down to about $3 \times 10^{-19}$ at 65536 s. The nearly flat behaviour of the FFS between 1s and 10s remains unexplained and seems uncorrelated with both room temperature fluctuations and optical power fluctuations. An additional control loop for stabilizing the power incident on the photodiode didn't modify the measured FFS or the phase noise.

To evaluate the accuracy of optical to microwave generation we have used the standard technique of deducing the frequency offset from the timing drift (top of fig. 3) [16]. The data shows a conversion accuracy of $2 \times 10^{-20}$, which is compatible with zero within the error bars defined by the FFS shown in fig. 3. Our result sets a new stringent limit, nearly two orders of magnitude better than previously reported on TSOFC and FOFC [16].

In conclusion, we have used two FOFC-based optical-to-microwave division frequency synthesizers referenced to a common optically source to create 11.55 GHz microwave signals with an unprecedented residual frequency stability of $1.6 \times 10^{-16}$ at 1 s. The residual single-sideband phase noise at a 1 Hz offset from the 11.55 GHz carrier is measured at −111 dBc/Hz, limited by the readout system noise floor. Long term stability and accuracy down to $3\times10^{-19}$ at 65536 s was also demonstrated from a set of 3 days continuous measurement.

**Figure captions**

Figure 1: Set-up schematic (OADM: Optical Add/Drop Module, AOM: Acousto Optic Modulator, DVM: Digital Voltage Multimeter, DDS: Direct Digital Synthesizer, USL: Ultra Stable Laser, PC: Polarization Controller, FFT: Fast Fourier Transform analyzer).

Figure 2: Single side band phase noise power spectral density of the difference of the two 11.55 GHz microwave signals (solid black line), measurement floor (photodetection+amplification) (dashed grey line) (200 averages/decade, DC coupling) and .the estimated cryogenic sapphire oscillator phase noise (dashed black line).

Figure 3: Top plot: relative timing jitter between the two microwave signals at 11 11.55 GHz over 72h (26300, 10s samples). The measurement is performed by sampling the voltage DC output of the microwave mixer (measurement bandwidth 5 Hz). Bottom plot : residual fractional frequency stability of the generated microwave signal scaled to one system (overlapping Allan deviation, 263000 1s samples).

Figure 1

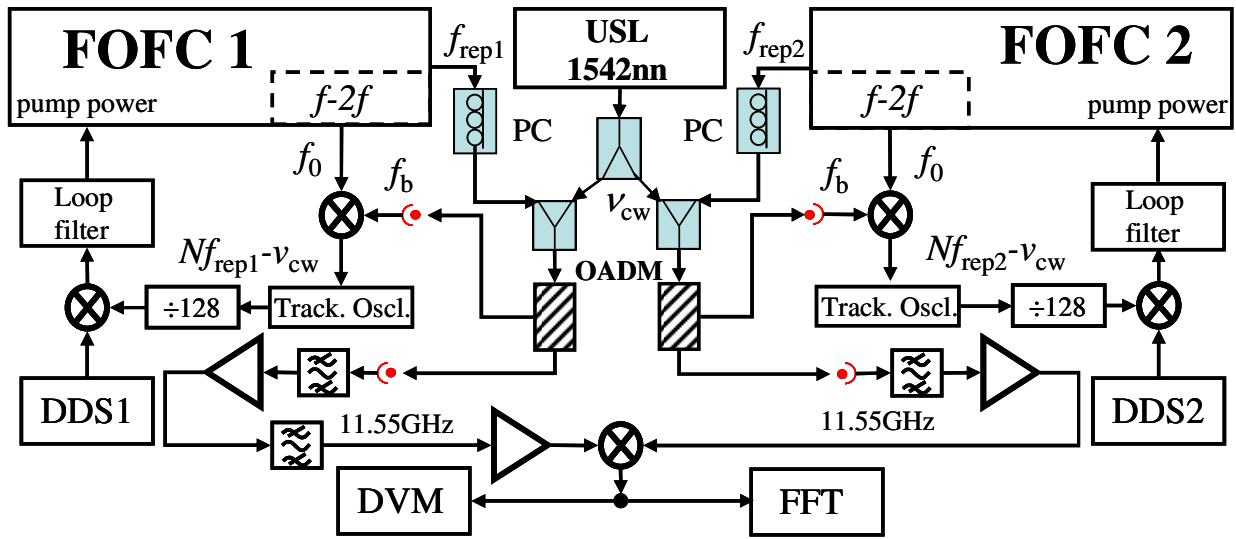

Figure 2

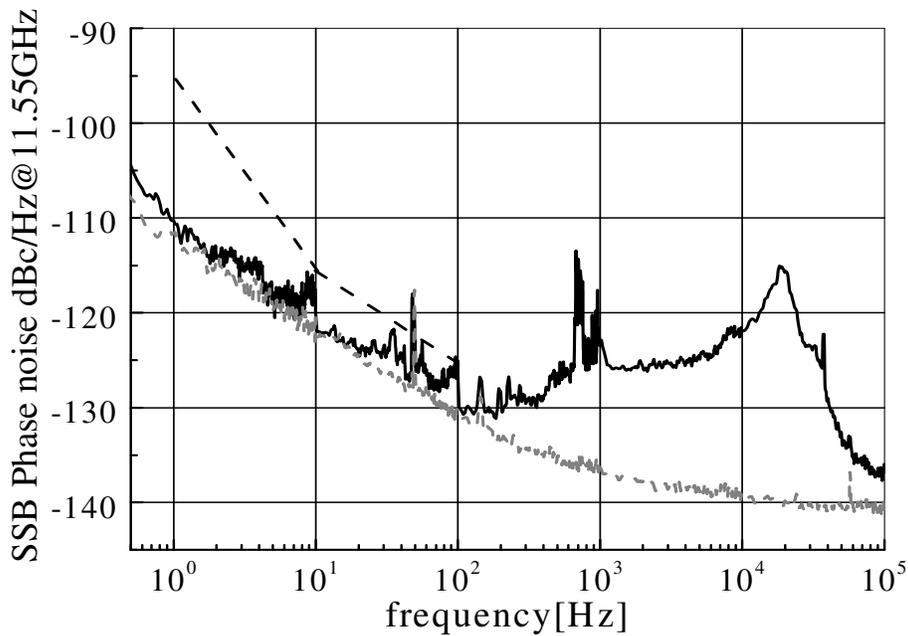

Figure 3

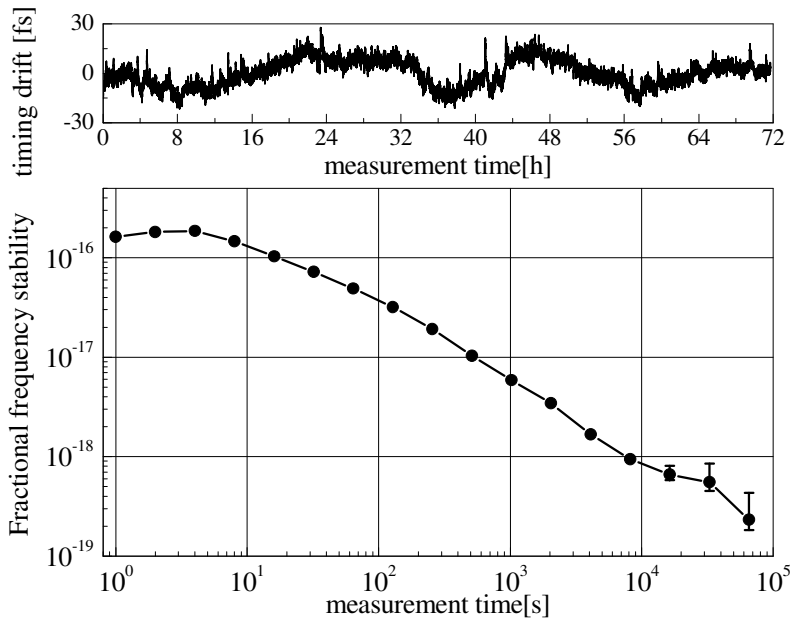